\documentclass[english,final]{article}

\usepackage{latexsym,babel,epsfig,color,amssymb,amsmath}
\usepackage{lineno,fullpage}

\newtheorem{theorem}{Theorem}
\newtheorem{corollary}[theorem]{Corollary}

\newtheorem{lemma}[theorem]{Lemma}
\newtheorem{remark}{Remark}
\newtheorem{fact}[theorem]{Fact}
 
\newtheorem{problem}{Problem} 
\newenvironment{proof}{{\it Proof:\/}}{\hfill$\Box$\vskip 0.1in}

\newcommand{\smallspace}{\;}
\newcommand{\nth}[1]{\ensuremath{{#1}^{\textup{th}}}}

\newcommand{\tree}{\mathcal{T}}
\newcommand{\opti}[1]{\delta(v_0,#1)}

\title{Quantum query complexity of some graph problems\thanks{This
    paper subsumes the three manuscripts
    ``A quantum algorithm for finding the minimum''
    (quant-ph/960714), ``Quantum algorithms for lowest weight paths and
    spanning trees in complete graphs'' (quant-ph/0303131), and ``Quantum
    query complexity of graph connectivity'' (quant-ph/0303169), and
    includes several other previously unpublished results.
    We are grateful to Yaohui Lei for his permission to include 
    results presented in quant-ph/0303169 in this paper.}}

\author{
Christoph D\"urr\thanks{%
  Laboratoire de Recherche en Informatique, 
  UMR 8623, Universit\'e Paris-Sud,
  91405 Orsay, France. durr@lri.fr. Research partially supported by
  the EU fifth framework program RESQ IST-2001-37559, and RAND-APX
  IST-1999-14036, by CNRS/STIC 01N80/0502 grant, by ACI Cryptologie
  CR/02 02 0040 grant of the French Research Ministry.}  \and
Mark Heiligman\thanks{%
  Advanced Research and Development Activity, Suite 6644, National
  Security Agency, 9800 Savage Road, Fort Meade, Maryland 20755, USA.
  miheili@nsa.gov.}  \and
Peter H\o{}yer\thanks{%
  Dept. of Computer Science, Univ. of Calgary, Alberta, Canada.
  hoyer@cpsc.ucalgary.ca.  Supported in part by the Alberta Ingenuity
  Fund and the Pacific Institute for the Mathematical Sciences.}  \and
Mehdi Mhalla\thanks{%
  Laboratoire Leibniz, Institut IMAG, Grenoble, France.
  Mehdi.Mhalla@imag.fr.}
}

\date{January 15, 2004}

\begin{document}

\maketitle

\begin{abstract} 
Quantum algorithms for graph problems are considered, both in the
adjacency matrix model and in an adjacency list-like array model.  We
give almost tight lower and upper bounds for the bounded error quantum
query complexity of \textsc{Connectivity}, \textsc{Strong
Connectivity}, \textsc{Minimum Spanning Tree}, and \textsc{Single
Source Shortest Paths}.  For example we show that the query complexity
of \textsc{Minimum Spanning Tree} is in $\Theta(n^{3/2})$ in the
matrix model and in $\Theta(\sqrt{nm})$ in the array model, while the
complexity of \textsc{Connectivity} is also in $\Theta(n^{3/2})$ in
the matrix model, but in $\Theta(n)$ in the array model.  The upper
bounds utilize search procedures for finding minima of functions
under various conditions.
\end{abstract}

\paragraph*{Keywords:} graph theory, quantum algorithm, lower bound,
  connectivity, minimum spanning tree, single source shortest paths

\section{Introduction}
\label{sec:intro}
  
  A primary goal of the theory of quantum complexity is to determine
  when quantum computers may offer a computational speed-up over
  classical computers.  Today there are only a few results which give
  a polynomial time quantum algorithm for some problem for which no
  classical polynomial time solution is known.  We are interested in
  studying the potentialities for speed-up for problems for which
  there already are efficient classical algorithms.  Basic graphs
  problems are interesting candidates.
  
  We study the query complexity of these problems; meaning the minimal
  number of queries to the graph required for solving the
  problem. Throughout this paper, the symbol $[n]$ denotes the set
  $[0..n-1]$.  We consider two query models for directed graphs:
  \begin{description} \item[The adjacency matrix model,] where the
  graph is given as the adjacency matrix $M\in\{0,1\}^{ n\times n} $,
  with $M_{ij}=1$ if and only if $(v_i,v_j) \in E$.

  \item[The adjacency array model,] where we are given the out-degrees
    of the vertices $d^+_1,\ldots,d^+_n$ and for every vertex $u$ an
    array with its neighbors $f_i : [d^+_i] \rightarrow [n]$.  So
    $f_i(j)$ returns the \nth{j} neighbor of vertex $i$, according to
    some arbitrary but fixed numbering of the outgoing edges of~$i$.
    In this paper the upper bounds for this model are all at least
    $n$, so we assume henceforth that the degrees are given as part of
    the input and we account only queries to the arrays $f_i$.  In
    addition the arrays satisfy the \emph{simple graph} promise
    \begin{linenomath}\[
      \forall i\in [n],j,j'\in [k],j\neq j' :
      f_i(j)\neq f_i(j')
    \]\end{linenomath}
    ensuring the graph is not a multigraph, i.e.\ does not 
    have multiple edges between any two vertices.  
  \end{description} 

  For undirected graphs we require an additional promise on the input,
  namely that $M$ is symmetric in the matrix model, and for the array
  model that $\forall i,i'\in [n]$ if $\exists j\in[k]: f_i(j)=i'$
  then $\exists j'\in[k]:f_{i'}(j')=i$.  Note that in the matrix model
  this symmetry assumption does not make undirected graph problems,
  promise problems since we may assume that the input is upper
  triangular.
  
  Weighted graphs are encoded by a weight matrix, where for
  convenience we set $M_{ij}=\infty$ if $(v_i,v_j)\not\in E$.  In the
  adjacency array model, the graph is encoded by a sequence of
  functions $f_i : [d^+_i] \rightarrow [n]\times \mathbb N$, such that
  if $f_i(j)=(i',w)$ then there is an edge $(v_i,v_{i'})$ and it has
  weight $w$.
  
  We emphasize that the array model is different {from} the standard
  list model.  In the latter, we have access to the neighbors of a
  given vertex only as a list, and thus querying the \nth{i} neighbor
  requires $i$ accesses to the list.  This is also true on a quantum
  computer, so its speedup is quite restricted.
  
  Many other query models are of course possible, for example we could
  be given an array of edges $f:[m]\rightarrow [n]\times [n]$, or an
  ordered array (which is up to $O(n)$ preprocessing the same as the
  adjacency array model).  For simplicity, we use the array model as
  presented above.

  For the quantum query complexity of general monotone graph
  properties, a lower bound of $\Omega(\sqrt n)$ is known in the
  matrix model, as shown by Buhrman, Cleve, de Wolf and
  Zalka~\cite{BCWZ}.\footnote{In a previous version of this paper, we
  said that Buhrman et al.{} conjectured $\Omega(n)$ for
  \textsc{Connectivity}, and since their conjecture concerns arbitrary
  monotone graph properties in general, we gave a false impression of
  improving their result.  We apologize.}  We are not aware of any
  quantum nor classical lower bounds in the array model.

        In this paper we show that the quantum query complexity of
  \textsc{Connectivity} is $\Theta(n^{3/2})$ in the matrix model and
  $\Theta(n)$ in the array model.  The classical randomized query
  complexity of \textsc{Connectivity} in the matrix model is
  $\Omega(n^2)$ by a sensitivity argument: Distinguishing the graph
  consisting of two length $n/2$ paths {from} the graph consisting of
  those two paths, plus an additional edge connecting them,
  $\Omega(n^2)$ queries are required.
  
  We study the complexity of three other problems. In \textsc{Strong
    Connectivity} we are given a directed graph and have to decide if
  there is a directed path between any pair of vertices.  In
  \textsc{Minimum Spanning Tree} we are given a weighted graph and
  have to compute a spanning tree with minimal total edge weight.  In
  \textsc{Single Source Shortest Paths} we have to compute the
  shortest paths according to the total edge weight {from} a given
  source vertex to every other vertex.  The quantum query complexity
  of these three problems is $\Omega(n^{3/2})$ in the matrix model and
  $\Omega(\sqrt{nm})$ in the array model.  We give almost tight upper
  bounds.

\begin{table}[htb]
\begin{center}
\begin{tabular}{llll} 
problem
                      & matrix model & array model \\ \hline
minimum spanning tree & $\Theta(n^{3/2})$ 
                              & $\Theta(\sqrt{nm})$\\ 
connectivity          & $\Theta(n^{3/2})$ 
                              & $\Theta(n)$ \\
strong connectivity   & $\Theta(n^{3/2})$
                              & $\Omega(\sqrt{nm})$, $O(\sqrt{nm\log n})$ \\
single src. short.\ paths 
                      &$\Omega(n^{3/2})$, $O(n^{3/2}\log^{2} n)$ 
                              & $\Omega(\sqrt{nm})$, 
                                $O(\sqrt{nm}\log^{2} n)$
\end{tabular}
\end{center}
\caption{Quantum query complexity of some graph problems}
\label{tab:results}
\end{table}
  
  We note that for graphs with a large number of edges
  $(m=\Theta(n^2))$, the complexities are (almost) the same in the
  matrix and array model for all problems but \textsc{Connectivity}.
  However, the models still differ in that case.  For example the test
  $(u,v)\in E$ costs a single query in the matrix model and
  $O(\sqrt{\min\{d^+_u,d^+_v\}})$ queries in the array model since we
  do not assume any order on the arrays $f_u$ and $f_v$.

  The time complexities of the algorithms are the same as the query
  complexities up to log-factors.  The algorithms given for
  connectivity and strong connectivity can be altered to also output
  the (strongly) connected components without increasing the
  asymptotic complexity.  The space requirement is $O(\log n)$ qubits
  and $O(n\log n)$ classical bits.  If we constraint the space (both
  classical and quantum) to $O(\log n)$ qubits, the problems may be
  solved by random walks.  Quantum random walks has been the subject
  of several papers~\cite{AAKV01,CCDFGS02,K03}, in particular for the
  \textsc{$st$-Connectivity} problem~\cite{Wa01}.
  
  Other work on the query complexity of graphs problems has been done
  independently of us.  For example \cite{Frei04} shows that testing
  whether a given graph is bi-partite needs $\Omega(n^{3/2})$
  queries in the matrix model.  Note that a graph is bi-partite iff it
  does not contain an odd cycle.  For the array model a lower bound
  can be constructed from our lower bound for connectivity, showing
  that $\Omega(\sqrt{nm})$ queries are necessary for any bounded error
  quantum algorithm which distinguishes a single even cycle from two
  disjoint odd cycles.  Matching upper bounds follow from our
  connectivity algorithms: simply construct a spanning forest of the
  graph.  Then color alternating black-white the nodes of each tree.
  And finally use the quantum search procedure to find an edge with
  endpoints of the same color.  Such an edge creates an odd cycle, and
  exists iff the graph contains an odd cycle.
  
\section{Tools used in this paper}
\label{sec:tools}

We use two fundamental tools.  The first is amplitude
amplification~\cite{BH,BHMT}, which we use when proving the upper
bounds, the second is Ambainis' lower bound technique~\cite{AM02}.

Amplitude amplification is a generalization of Lov Grover's search
algorithm~\cite{Gro96}.  Since it is the most important tool used in
our algorithms, we restate the exact results we require.  We are given
a boolean function $F$ defined on a domain of size $n$.  The function
is given as a black~box so that the only way we can obtain information
about $F$ is via evaluating $F$ on elements in the domain.  The search
problem considered by Grover is to find an element $x$ for which
$F(x)=1$, provided one exists.  We say that $x$ is a \emph{solution}
to the search problem, and that $x$ is \emph{good}.  We use three
generalizations of the search algorithm---all of which we refer to as
``the search algorithm''.

\begin{itemize}

\item If there are $t$ elements mapped to~$1$ under $F$, with $t>0$,
the search algorithm returns a solution after an expected number of at
most $\frac9{10}\sqrt{n/t}$ queries to $F$.  The output of the
algorithm is chosen uniformly at random among the $t$ solutions.  The
algorithm does not require prior knowledge of~$t$~\cite{BBHT}.

\item A second version uses $O(\sqrt n)$ queries to $F$ in the worst
case and outputs a solution with probability at least a constant,
provided there is one~\cite{BBHT}.

\item A third version uses $O(\sqrt{n \log 1/\epsilon})$ queries to
$F$ and finds a solution with probability at least $1-\epsilon$,
provided there is one~\cite{BCWZ}.

\end{itemize}
  
We note that for very sparse graphs given in the adjacency matrix
model, it may for some applications be efficient to initially learn
all entries of the matrix by reiterating the first version of the
search algorithm, for instance as formalized in
Fact~\ref{fact:learnitall}.

\begin{fact}\label{fact:learnitall}
Let $k$ be given.  There is a quantum algorithm that takes as input an
$n \times n$ boolean matrix $M$, uses $O(n \sqrt{k})$ queries to $M$,
and outputs a set $S$ of $1$-entries of $M$ so that with probability
at least $\frac12$, $S$ is of cardinality at least~$k$ or contains all
1-entries of $M$ in case $M$ has less than $k$ 1-entries.
\end{fact}

Our lower bounds uses a technique introduced by Andris Ambainis.

\begin{theorem}[{\cite[theorem 6]{AM02}}]
Let $L\subseteq \{0,1\}^*$ be a decision problem. Let $X\subseteq L$
be a set of positive instances and $Y\subseteq \overline L$ a set of
negative instances.  Let $R\subseteq X\times Y$ be a relation between
instances of same size.  Let values $ m,m',\ell_{x,i} $ and $
\ell'_{y,i} $ for $x,y\in\{0,1\}^n$ and $x\in X$, $y\in Y$, $i\in[n]$
be so that
\begin{itemize}
 \item for every $x\in X$ there are at least $m$ different $y\in Y$
      in relation with $ x $,
 \item for every $y\in Y$ there are at least $m'$ different $x\in
      X$ in relation with $y$,
 \item for every $x\in X$ and $i\in[n]$ there are at most $\ell_{x,i}
      $ different $y\in Y$ in relation with $x$ which differ {from} $x$
      at entry $i$,
 \item for every $y\in Y$ and $i\in [n]$ there are at most
      $\ell'_{y,i} $ different $x\in X$ in relation with $y$ which differ
      {from} $y$ at entry $i$.
\end{itemize}
Then the quantum query complexity of $L$ is $\Omega
(\sqrt{mm'/\ell_{\max }}) $ where $ \ell_{\max }$ is the maximum of
$\ell_{x,i}\ell'_{y,i}$ subject to $x R y$ and $x_i \neq y_i$.
\end{theorem}

\section{Minima finding}
\label{sec:minima}

Many graph problems are optimization problems, as are finding a
minimum spanning tree, single source shortest paths, and largest
connected components.  Most quantum algorithms for such optimization
problems utilize the search algorithm discussed above.  A~very basic
and abstract optimization problem is as follows.  Suppose we are given
a function $f$ defined on a domain of size $n$, and we want to find an
index $i$ so that $f(i)$ is a minimum in the image of~$f$.  This
minimization problem was considered in~\cite{DH} which gives an
optimal quantum algorithm that uses $O(\sqrt{n})$ queries to $f$ and
finds such an $i$ with constant probability.  It is very simple to
analyze, and we present it now, since it will make the rest of the
section easier to understand.

\begin{enumerate}
  \item Initially let $j\in[N]$ be an index chosen uniformly at random.
  \item Repeat forever
    \begin{enumerate}
      \item Find an index $i\in[N]$ such that $f(i)<f(j)$.
      \item Set $j:=i$.
    \end{enumerate}
\end{enumerate}
\begin{theorem}[\cite{DH}]
  The expected number of queries to $f$, until $j$ contains the index
  of a minimum in the image of~$f$ is $O(\sqrt{N})$.
\end{theorem}
\begin{proof}
  Without loss of generality assume that $f$ is injective. Now every
  index $j\in [N]$ has a \emph{rank}, which we define as the number of
  indexes $i$ such that $f(i)\leq f(j)$.

  Let $p_r$ be the probability that at some moment of the algorithm,
  the index $j$ will have rank $r$.  We claim that $p_r=1/r$.
  Consider the first moment when $j$ will have rank less or equal $r$.
  This moment will happen with probability $1$.  At that moment,
  because of the uniform choice in step~1 and since step~2(a) uses the
  quantum search procedure, $j$ will be uniformly chosen among all
  indexes with rank less or equal $r$, so $p_r=1/r$.

  If $j$ has rank $r$, the search procedure of step~2(a) will require
  $c\sqrt{N/(r-1)}$ expected number of queries, for some constant $c$.
  Therefore the total number of queries until $j$ contains the
  solution is
\begin{linenomath}\[
  \sum_{r=2}^N p_r c\sqrt{N/(r-1)} 
  < c \sqrt{N} \sum_{r=1}^{N-1} r^{-3/2}
  < c \sqrt{N} \left(1+\int_{r=1}^{N-1} r^{-3/2} \textbf{d}r\right)
    =O(\sqrt{N})
\]\end{linenomath}
\end{proof}

Let $c'\sqrt{N}$ be the expected number of queries to $f$ until $j$
contains the index of a minima.  Stopping the algorithm after $2c'$
queries gives a quantum algorithm with error probability upper bounded
by $1/2$.

\begin{figure}[htb]
\begin{center}
\begin{tabular}{@{}ccc}
\epsfig{file=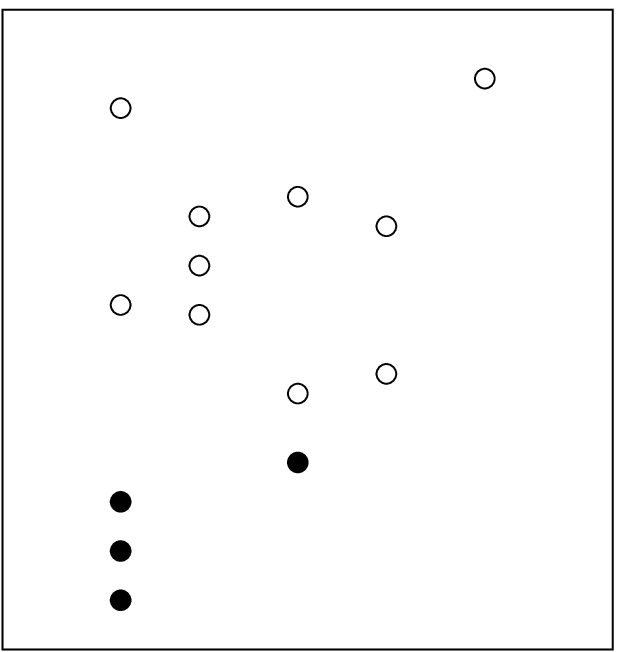,width=3cm}&
\epsfig{file=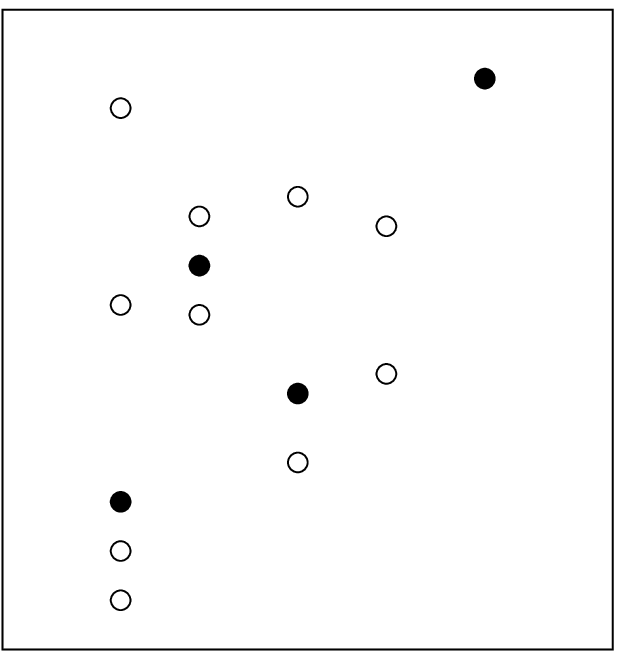,width=3cm}&
\epsfig{file=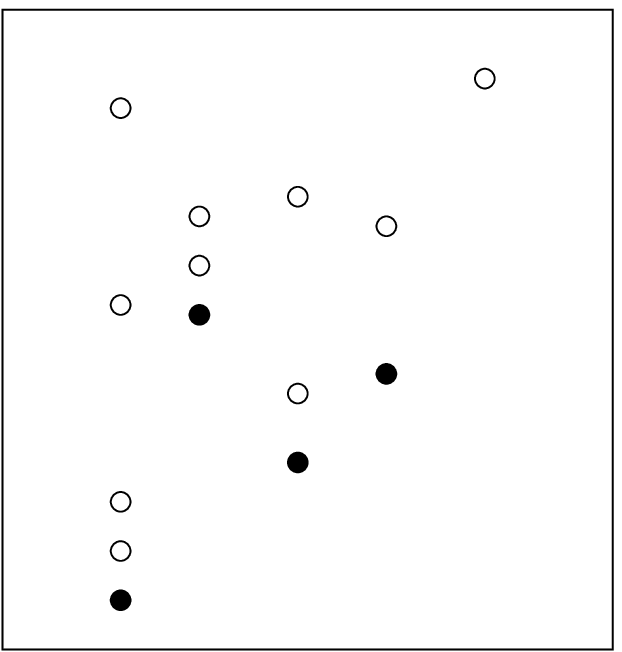,width=3cm}\\
4 points of min.\ value &
4 points of diff.\ type&
4 points of min.\ value \\[0cm]
&&and diff.\ type
\end{tabular}
\end{center}
\caption{Illustration of the three problems.  Each index $i$ is
illustrated by a point with horizontal coordinate $g(i)$ and vertical
coordinate $f(i)$.}
\label{fig:min}
\end{figure}

For the purposes of this paper, we require the following
generalizations of the minimum finding problem, illustrated in
Figure~\ref{fig:min}.
 
\begin{problem}[Find $d$ smallest values of a function]
\label{prob:dsmallest}
  Let $\mathbb N^*$ denote $\mathbb N \cup \{\infty\}$.  Given
  function $f:[N]\rightarrow \mathbb N^*$ and an integer $d\in [N]$,
  we wish to find $d$ distinct indexes mapping to smallest values,
  i.e.\ a subset $I \subseteq [N]$ of cardinality $d$ such that for
  any $j \in [N] \setminus I$ we have that $f(i) \leq f(j)$ for all $i
  \in I$.
\end{problem}

In the rest of this section, we assume $d\leq N/2$.  In the following
problem we are given a different function $g:[N]\rightarrow \mathbb
N$, such that $g(j)$ defines the \emph{type} of $j$.  Let $e=|\{g(j):
j\in [N]\}|$ be the number of different types.

\begin{problem}[Find $d$ elements of different type]
\label{prob:dtypes}
  Given function $g$ and an integer $d'$ we wish to find integer
  $d=\min\{d',e\}$ and a subset $I \subseteq [N]$ of cardinality
  $d$ such that $g(i) \neq g(i')$ for all distinct $i,i'\in I$.
\end{problem}

Now we present a generalization of both previous problems.

\begin{problem}[Find $d$ smallest values of different type]
\label{prob:dsmallesttypes}
  Given two functions $f,g$ and an integer $d'$ we wish to find
  integer $d=\min\{d',e\}$ and a subset $I \subseteq [N]$ of
  cardinality $d$ such that $g(i) \neq g(i')$ for all distinct
  $i,i'\in I$ and such that for all $j \in [N]\setminus I$ and $i \in
  I$, if $f(j) < f(i)$ then $f(i') \leq f(j)$ for some $i' \in I$ with
  $g(i')=g(j)$.  
\end{problem}

It is clear that Problems~\ref{prob:dsmallest} and~\ref{prob:dtypes}
are special cases of Problem~\ref{prob:dsmallesttypes}.  In this
section, we give an upper bound of $O(\sqrt{dN})$ for
Problem~\ref{prob:dsmallesttypes}.  In Section~\ref{sec:lower}, we
then show a lower bound of $\Omega(\sqrt{dN})$ for
Problems~\ref{prob:dsmallest} and~\ref{prob:dtypes}, implying that all
three problems are of complexity $\Theta(\sqrt{dN})$.  We prove the
upper bound by a simple greedy algorithm.  Consider a subset $I
\subseteq [N]$ of $d$ indices of different types.  We say an index $j
\in [N]$ is \emph{good for $I$} if
\begin{enumerate}
\item either $g(j)=g(i)$ and $f(j)<f(i)$ for some $i\in I$,
\item or $g(j) \not\in g(I)$ and $f(j) < f(i)$ for some $i\in I$.
\end{enumerate}
In the former case we say $j$ is a good index of \emph{known type},
in the latter that $j$ is a good index of \emph{unknown type}.  In
each iteration of the greedy algorithm, we find a good index~$j$ by
the search algorithm and then improve $I$ by replacing some index in
$I$ by~$j$.

\begin{enumerate}
\item Initially, let $I=\{N+1,\ldots, N+ d'\}$ be a set of artificial
      indices of unique different types and unique maximal value.
\item Repeat forever
\begin{enumerate}
\item \label{item:startloop} 
      Let $t$ denote the number of good elements for $I$.
      (Note: This step is not required, but only included for the 
      purpose of simplifying the analysis of the algorithm.)
\item \label{item:pickx} Use the first version of the search
      algorithm to find a good element $j \in [N]$ for~$I$.
\item Set $I = \textup{improve}(I,j)$ where we improve $I$ by
      replacing with $j$ the element in $I$ that has the same type as
      $j$ if $j$ is of known type, and by replacing with $j$ some
      element in $I$ with largest $f$-value if $j$ is of unknown type.
\end{enumerate}
\end{enumerate}

The next lemma shows we only need an expected number of $O(d)$
iterations of the main loop to eliminate a constant fraction of the
remaining good elements.

\begin{lemma}
Let $I \subseteq [N]$ be any subset of $d'$ indices of different types
with $t>0$ good elements of $e$ types.  After an expected number of
$O(d)$ iterations of the main loop there are at most $\frac34t$
good elements for~$I$.  Here $d= \min\{d', e\}$.
\end{lemma}

\begin{proof}
For notational simplicity assume $f$ is injective.  Set $I_0=I$ and
let $T_0=T$ be the set of good elements for~$I$.  Let $T_j$ denote the
set of good elements after $j$ iterations of the main loop, for $j>0$.
Similarly, let $I_j$ denote the selected index-set after $j$
iterations, for $j > 0$.
Set $t_k=|T_k|$.  In particular $I_0=I$ and $t_0=t$.  Let
$y_{\textup{mid}}$ denote the \nth{\lfloor t/2 \rfloor} smallest of
the $t$ elements according to~$f$.  For any subset $S \subseteq
[N+d']$, let $\textup{low}(S)$ denote the number of elements in $S$
that are no bigger than $y_{\textup{mid}}$ according to~$f$.

Note that initially
\begin{itemize}
\item $\textup{low}(T_0) = \lfloor t/2 \rfloor$ and 
\item $\textup{low}(I_0) < d$.
\end{itemize}
By the nature of the greedy algorithm, $\textup{low}(T_{k+1}) \leq
\textup{low}(T_{k})$ and $\textup{low}(I_{k+1}) \geq
\textup{low}(I_{k})$ for any $k \geq 0$.  Note that
\begin{itemize}
\item if $\textup{low}(T_{k}) < \frac{t}{4}$, then we have
eliminated at least a fraction of $\frac{1}{4}$ of the initially $t$
good elements for $I$, and similarly,
\item if $\textup{low}(I_{k}) = d$, then we have eliminated at least a
fraction of $\frac{1}{2}$ of the initially $t$ good elements
for~$I$.
\end{itemize}
We claim that in each iteration of the main loop, as long as
$\textup{low}(T_k) \geq \frac{t}{4}$, with probability at least
$\frac{1}{32}$, at least one of the following two events happens
\begin{itemize}
\item $\textup{low}(T_{k+1}) \leq 
         \textup{low}(T_{k}) \left(1-\frac{1}{32d}\right)$, 
\item $\textup{low}(I_{k+1}) = \textup{low}(I_{k}) +1$.
\end{itemize}

Assume $\textup{low}(T_k) \geq \frac{t}{4}$, since otherwise we are
done.  Consider the element $j$ picked in Step~\ref{item:pickx}.
First suppose the majority of the $\textup{low}(T_k)$ indices are of
unknown type with respect to~$I_k$.  Then, with probability at least
$\frac{1}{8}$, index $j$ is among these, in which case
$\textup{low}(I_{k+1}) = \textup{low}(I_{k})+1$.

Now suppose the majority of the $\textup{low}(T_k)$ indices are of
known type with respect to~$I_k$.  Then, with probability at least
$\frac{1}{8}$, index $j$ is among these.  Conditioned on this happens,
with probability at least $\frac{1}{2}$, there are at least
$\frac{\textup{low}(T_k)}{4d}$ good elements for $I_k$ of the same
type as~$j$.  With probability at least $\frac{1}{2}$, at least half
of these are not good for $I_{k+1}$.  Thus, with probability at least
$\frac{1}{32}$, we have eliminated at least $\frac{t}{32d}$ of the
remaining elements in~$T_j$.  

This proves the claim.  It follows that after an expected number of
$O(d)$ iterations of the main loop, we have eliminated at least a
fraction of $\frac{1}{4}$ of the initially $t$ good elements.
\end{proof}

The above lemma implies that, for $t > 2d$, after an expected number
of $O(d \sqrt{N/t}\smallspace)$ applications of function~$f$, the
number of good elements is at most $\frac{t}{2}$.  
Hence, for any $t> 2d$, the expected number
applications of function~$f$ required till we have that $t \leq 2d$ for
the first time is in the order of 
\begin{linenomath}\begin{equation*}
d\Big(\sqrt{\frac{N}{d}} + \sqrt{\frac{N}{2d}} + \sqrt{\frac{N}{4d}} + 
\sqrt{\frac{N}{8d}} + \cdots\Big) \in O(\sqrt{dN}).
\end{equation*}\end{linenomath}
Once $t \leq 2d$ for the first time, the expected number of
applications of $f$ required before $t=0$ for the first time is in the
order of $\sum_{j=1}^{2d} \sqrt{{N}/{j}}$ which is in
$O(\sqrt{dN}\smallspace)$.

\begin{corollary}
In the greedy algorithm given above, after an expected number of
$O(\sqrt{dN})$ applications of function $f$, there are no good
elements for~$I$, that is, $t=0$.
\end{corollary}

The next theorem follows immediately.

\begin{theorem}
  The problem \textsc{Find $d$ Smallest Values of Different Type} has
  bounded error quantum query complexity $O(\sqrt{dN})$.
\end{theorem}

Nayak and Wu give in~\cite{NW} a bounded error quantum algorithm that
given a function $f:[N]\rightarrow \mathbb N$ and two integers $d$ and
$\Delta$, outputs an index $i$ such that the rank of $f(i)$ is between
$d-\Delta$ and $d+\Delta$.  The query complexity of their algorithm is
$O(M \log M \log\log M)$ where $M=\sqrt{N/\Delta} +
\sqrt{d(N-d)}/\Delta$.  Setting $\Delta=\frac12$ it would find the
\nth{d} smallest element with $O(\sqrt{dN}\log N \log \log N)$
queries. Nayak~\cite{N} later improved this algorithm to
$O(\sqrt{dN})$, matching the lower bound given in~\cite{NW}.  His
method is different from ours.

\begin{remark}\label{remark:min} 
  The algorithm above uses $c\sqrt{dN}$ queries for some constant $c$
  and outputs the solution with probability at least $1/2$.  In order
  to reduce the error probability to $1/2^k$ one could run the
  algorithm $k$ times and among the $dk$ resulting indices, output the
  $d$ smallest values of different type.  However starting each run
  with randomly chosen points of different type regardless of the
  previous outcome, is a waste of information.  So it is much more
  clever to run the algorithm only once and stop it after
  $kc\sqrt{dN}$ queries.
\end{remark}

\section{Minimum Spanning Tree}
\label{sec:mst}

In this section we consider undirected graphs with weighted edges.  In
\textsc{Minimum Spanning Tree} we wish to compute a cycle-free edge
set of maximal cardinality that has minimum total weight.  To be
precise if the graph is not connected this is actually a {spanning
forest}.

Classically, there are a number of different approaches to finding
minimum spanning trees efficiently, including the algorithms of
Bor\r{u}vka~\cite{Boruvka,NMN}, Kruskal~\cite{Kruskal}, and
Prim~\cite{Prim}.  To construct an efficient quantum algorithm, we use
Bor\r{u}vka's algorithm since it is of a highly parallel nature.  This
allows us to use the minima finding algorithms given in
Section~\ref{sec:minima}.

Bor\r{u}vka's algorithm consists of at most $\log n$ iterations.  In
brief, initially it starts with a collection of $n$ spanning trees,
each tree containing a single vertex.  In each iteration, it finds a
minimum weight edge out of each tree in the collection, adds the edges
to the trees, and merges them into larger and fewer trees.  After at
most $\log n$ iterations, there is only one tree left, which is a
minimum spanning tree.  The correctness of Bor\r{u}vka's algorithm rests
on the following simple fact about spanning trees.

\begin{fact} 
Let $U\subset V$ be a set of vertices of a connected graph $G=(V,E)$
and let $e$ be a minimum weight edge of $(U\times \overline U)\cap E$.
Then there is a minimum spanning tree containing~$e$.
\end{fact}

In our quantum version of Bor\r{u}vka's algorithm, we make a few
adjustments to keep the overall error probability small without
sacrificing in the number of queries.  We adjust it slightly so that
the \nth{\ell} iteration errs with probability at most
$\frac{1}{2^{\ell+2}}$, ensuring that the overall error is at
most~$\frac{1}{4}$.  This increases the cost of the \nth{\ell}
iteration by a factor of $\ell$, but since the cost of the first few
iterations dominates, this is asymptotically negligible.  The details
follow.

\begin{enumerate} 
\item Let $T_1, T_2, \ldots, T_k$ be a spanning forest.
      Initially, $k=n$ and each tree $T_j$ contains a single vertex.
\item Set $\ell=0$.
\item Repeat until there is only a single spanning tree (i.e., $k=1$).
\begin{enumerate}
\item Increment $\ell$.
\item \label{item:step:MST}
Find edges $e_1, e_2, \ldots, e_k$ satisfying that $e_j$ is a
minimum weight edge leaving $T_j$.  
Interrupt when the total number of queries is
$(\ell+2)c\sqrt{km}$ for some appropriate constant~$c$.
\item Add the edges $e'_j$ to the trees, merging them into larger
trees.
\end{enumerate}
\item Return the spanning tree~$T_1$.
\end{enumerate}

To find the minimum edges $e_1,\ldots,e_k$ in
Step~\ref{item:step:MST}, we use the following functions.  In the
array model, any edge $(u,v)$ is coded twice, $u$ appears as neighbor
of $v$, but $v$ also appears as neighbor of $u$.  Enumerate the
directed edges {from} $0$ to $2m-1$.  Let $f:[2m]\rightarrow\mathbb
N^*$ denote the function that maps every directed edge $(u,v)$ to its
weight if $u$ and $v$ belong to different trees of the current
spanning forest and to $\infty$ otherwise.  Let $g:[2m]\rightarrow
[k]$ denote the function that maps every directed edge $(u,v)$ to the
index $j$ of the tree $T_j$ containing~$u$.  We then apply the
algorithm for \textsc{Finding $k$ smallest values of different type},
interrupting it after $(\ell+2)c\sqrt{km}$ queries to obtain an error
probability at most $1/2^{\ell+2}$, see Remark~\ref{remark:min}.

\begin{theorem}
Given an undirected graph with weighted edges, the algorithm above
outputs a spanning tree that is minimum with probability at
least~$\frac14$.  The algorithm uses $O(\sqrt{nm})$ queries in the
array model and $O(n^{3/2})$ queries in the matrix model.
\end{theorem}

\begin{proof}
To simplify the proof, consider the matrix model an instance of
the array model with $m=n(n-1)$ edges.

At the beginning of the \nth{\ell} iteration of the main loop, the
number of trees $k$ is at most $n/2^{\ell-1}$, and thus it uses at
most $(\ell+2)c \sqrt{nm/2^{\ell-1}}$ queries.  Summing over all
iterations, the total number of queries is at most $\sum_{\ell \geq
1}(\ell+2)c \sqrt{nm/2^{\ell-1}}$, which is in $O(\sqrt{nm})$.

The \nth{\ell} iteration introduces an error with probability at most
$\frac{1}{2^{\ell+2}}$.  The overall error probability is thus upper
bounded by $\sum_{\ell \geq 1} \frac{1}{2^{\ell+2}} \leq \frac14$.
\end{proof}

For our algorithms, a $O(\log n)$ factor applies when considering the
bit computational model, rather than the algebraic computational
model.  Apart from that, the time complexity is the same as the query
complexity by using an appropriate data structure.  Each vertex holds
a pointer to another vertex in the same component, except for a unique
vertex per component that holds a null pointer.  This vertex is called
the canonical representative of the component.  To decide if two
vertices are in the same component, we need only to determine the
canonical representative of each vertex by pointer chasing.  To merge
two components, we change the pointer of one of the canonical
representative to point to the other.

Using pointer chasing, the time complexity of the \nth{\ell} iteration
is a factor of~$\ell$ larger than its query complexity.  However, as
in the case of the error reduction, this is insignificant: $\sum_{\ell
\geq 1} \ell (\ell+2)c \sqrt{nm/2^{\ell-1}}$ is also in
$O(\sqrt{nm})$.  Thus, the time complexity is asymptotically the same
as the query complexity.

\section{Connectivity}
\label{sec:connectivity}

A~special case of \textsc{Minimum Spanning Tree} when all edge weights
are equal, is \textsc{Graph Connectivity}. The input is an
\emph{undirected} graph and the output is a spanning tree, provided
the graph is connected.  

For the matrix model, the algorithm for minimum spanning tree given in
the previous section implies an $O(n^{3/2})$ upper bound for graph
connectivity as well.  Below, we give a somewhat simpler and arguably
more natural quantum algorithm of query complexity $O(n^{3/2})$, which
is optimal by the lower bound given in Section~\ref{sec:lower} below.

For the array model, we give a quantum algorithm that uses only $O(n)$
queries.  Both algorithms start with a collection of $n$ connected
components, one for each vertex, and greedily construct a spanning
tree by repeatedly picking an edge that connects two of the
components.

\begin{theorem}\label{thm:adj-undirected-spanningtree}
Given the adjacency matrix $M$ of an undirected graph $G$, the
algorithm below outputs a spanning tree for~$G$ after an expected
number of $O(n^{3/2})$ queries to $M$, provided $G$ is connected, and
otherwise runs forever.
\end{theorem}

\begin{proof}
Consider the following algorithm.
\begin{enumerate} 
\item Initially the edge set $A$ is empty.
\item Repeat until $A$ connects the graph.
\begin{enumerate}
\item Search for a \emph{good} edge, i.e., an edge that connects two
different components in $A$, and add it to~$A$.  Use the version of
the search algorithm that returns a solution in expected $O(n^2/t)$
queries if there are $t>0$ good edges and otherwise runs forever.
\end{enumerate}
\item Return the edge set $A$.
\end{enumerate}

Suppose the graph is connected and consider the expected total number
of queries used by the algorithm.  There are exactly $n-1$ iterations
of the main loop.  The number of good edges is at least $k-1$ when $A$
consists of $k$ components, and thus the expected total number of
queries is in the order of $\sum_{k=2}^{n} \sqrt{n^2/(k-1)}$, which is
in $O(n^{3/2})$.
\end{proof}

When implementing the above algorithm, we maintain an appropriate data
structure containing information about the connected components in the
graph induced by $A$.  This introduces an additional $O(n\log n)$ term
in the running time of the algorithm which is negligible compared to
$O(n^{3/2})$.  We may choose to stop the algorithm after twice the
expected total number of queries, giving an $O(n^{3/2})$ query
algorithm with bounded one-sided error.

\subsection{The array model}

\begin{lemma}\label{lm:arr-undirected-preproc}
Given an undirected graph $G$ in the array model, we can in $O(n)$
classical queries construct a set of connected components $\{C_1,
\ldots, C_k\}$ for some integer $k$, so that for each component $C$,
its total degree $m_C = \sum_{i\in C}d_i$ is no more than $|C|^2$.
\end{lemma}

\begin{proof}
The algorithm is classical and is as follows.
\begin{enumerate} 
\item Initially the edge set $A$ is empty.
\item Let $S=V$ be the set of vertices not yet placed in some 
      component.
\item Let $k=0$ be the number of components constructed thus far.
\item While $S$ is non-empty
\begin{enumerate}
 \item \label{item:step:pickhighdegree}
  Take the vertex $v$ of highest degree in $S$ and set $D=\{v\}$.
 \item \label{item:step:addmore}
  Go through $v$'s list of neighbors one  by one, each time adding
 the neighbor $w$ to $D$ and the edge $(v,w)$ to $A$, until one of two
 events happens: (1) We reach the end of the list, or (2) we reach a
 neighbor $w$ already assigned to some component $C_j$ with $j\leq k$.
 \item In case (1), set $k=k+1$, $C_k=D$, and remove $D$ {from} $S$.  In
 case (2), add $D$ to $C_j$, and remove $D$ {from} $S$.
\end{enumerate}
\item Output $k$, $A$, and $C_1, C_2, \ldots, C_k$.
\end{enumerate}

The algorithm uses $n-k$ queries in total, one query for each vertex
but the first added to each component.  Edge set $A$ contains the
union of spanning trees of the components $C_1$ through~$C_k$.

To show correctness, let $v$ be the vertex chosen in
Step~\ref{item:step:pickhighdegree} and $d$ its degree.  Then $d\le
|C_j|$ for each components constructed so far, since the size of a
freshly created component is the degree of one of its vertices, which
by choice in Step~\ref{item:step:pickhighdegree} must be no less than
$d$, and components can only grow.  To show that the total degree of
every component $C_j$ is no more than $|C_j|^2$, consider the two
cases in Step~\ref{item:step:addmore}.

In case (1), $D$ is the set of $v$ and its neighbors, each neighbor
having degree no larger than~$d$, implying the total degree is at most
$d(d+1)$ which is strictly less than $(d+1)^2=|D|^2$.
In case (2), let $a$ be the size of the component $C_j$ to which $D$
is merged, and $b$ the size of $D$.  Then $b\leq d\leq a$.  The total
degree is no more than $a^2+bd$ which is strictly less than $(a+b)^2$.
\end{proof}

\begin{theorem}
Given an undirected graph $G$ in the array model, the algorithm below
outputs a spanning tree for~$G$ using an expected number of $O(n)$
queries, provided $G$ is connected, and otherwise runs forever.
\end{theorem}

\begin{proof}
Consider the following algorithm.
\begin{enumerate} 
\item Construct the edge set $A$ using the above lemma.
\item Repeat until $A$ connects the graph.
\begin{enumerate}
\item Pick a connected component $C$ in $A$ with smallest total
degree, i.e, a component minimizing $m_C = \sum_{i \in C} d_i$.
\item Search for an edge out of $C$, i.e., an edge that connects $C$
to some other component in $A$, and add it to~$A$.  Use the version of
the search algorithm that returns a solution in expected $O(\sqrt
{m_C})$ queries if there is at least one such edge and otherwise runs
forever.
\end{enumerate}
\item Return the edge set $A$.
\end{enumerate}

Suppose the graph is connected and consider the expected total number
of queries used by the algorithm.  

We first construct $k$ components, each component $C$ having total
degree $m_C$ at most $|C|^2$.  In each iteration of the main loop, we
pick the component with smallest total degree and search for an edge
out of~$C$.  The expected cost of finding such an edge is at most
$\alpha\sqrt{m_C}$ for some constant~$\alpha$.  We distribute this
cost evenly among each of the $m_C$ edge endpoints in $C$, each
endpoint paying $\alpha/\sqrt{m_C}$.

Fix an arbitrary edge endpoint.  Enumerate {from} $0$ up to at most
$\log m$ the successive components that were chosen by the algorithm
for a search and that contain this fixed edge endpoint.  Let $m_i$ be
the number of edge endpoint in the \nth{i} component.  Then $m_{i+1}
\geq 2m_i$.  The total cost assigned to our fixed edge endpoint is
upper bounded by
\begin{linenomath}\begin{equation*}
\sum_{i = 0}^{\log m} \frac{\alpha}{\sqrt{m_i}} \leq 
\sum_{i = 0}^{\log m} \frac{\alpha}{\sqrt{2^i m_0}} \leq 
\frac{4\alpha}{\sqrt{m_0}}.
\end{equation*}\end{linenomath}

Let $C$ be any of the $k$ components constructed in the first step.
The total cost assigned over all edge endpoints in $C$ is thus upper
bounded by ${4\alpha}{\sqrt{m_C}}$, which is at most $4\alpha|C|$.
Summing over all $k$ components, the total cost assigned in the main
loop is at most $4\alpha n$, which is linear in~$n$.
\end{proof}

\section{Strong Connectivity}
\label{sec:strongconnectivity}

We give two quantum algorithms for strong connectivity, first one for
the matrix model and then one for the array model.  The input is a
\emph{directed} graph and the output is a set of at most $2(n-1)$
edges that proves the graph is strongly connected, provided it is.  It
follows {from} the discussions below that such sets always exist.

\begin{theorem}\label{thm:adj-dir-spanningtree}
Given the adjacency matrix $M$ of a directed graph $G$ and a vertex
$v_0$, the algorithm below uses $O(\sqrt{nm\log n})$ queries to
$M$ and outputs a directed tree $A \subseteq E$ rooted at $v_0$.  With
probability at least $\frac{9}{10}$, $A$ spans all of $G$, provided
such a spanning tree exists.
\end{theorem}

\begin{proof}
Consider the following simple algorithm.
\begin{enumerate} 
\item Initially the edge set $A$ is empty.
\item Let $S=\{v_0\}$ be a set of reachable vertices, and $T=\{v_0\}$
  a stack of vertices to be processed.
\item While $T \neq \{\}$ do
\begin{enumerate}
\item Let $u$ be the top most vertex of stack~$T$.
\item Search for a neighbor $v$ of $u$ not in $S$.  Use the version of
      the search algorithm that uses $O(\sqrt{d^+_u\log n})$ queries  and
      outputs a solution with probability at least $1-\frac{1}{20n}$,
      provided one exists.
\item If succeed, add $(u,v)$ to $A$, add $v$ to $S$, and 
      push $v$ onto~$T$.
\item Otherwise, remove $u$ {from}~$T$.
\end{enumerate}
\item Return edge set~$A$.
\end{enumerate}

For any vertex $u$ let $b^+_u$ the out-degree in the tree $A$ produced
by the algorithm.  Then the total number of queries spent in finding
the $b^+_u$ neighbors of $u$ is in the order of $\sum_{t=1}^{b^+_u}
\sqrt{d^+_u/t} \sqrt{\log n}$ which is in $O(\sqrt{b^+_u d^+_u \log
n})$.  Summing over all vertices $u$ this gives
\begin{linenomath}\[
        \sum_{u\in V} \sqrt{b^+_u d^+_u \log n} 
        \le \sqrt{\sum_u b^+_u} \sqrt{\sum_u d^+_u} \sqrt{\log n}
        = O(\sqrt{nm\log n}).
\]\end{linenomath}
The first inequality follows from the general statement that the inner
product of two vectors is upper bounded by the product of their
${\ell}_2$-norms.  The second inequality uses the fact that a tree has
only $O(n)$ edges.  The algorithm spends in addition $O(\sum_u
\sqrt{d^+_u\log n})$ queries for the unsuccessful neighbor searches,
but this is dominated by the previous cost.

The overall error probability is upper bounded by $\frac{1}{10}$ since
each of the at most $2n-1$ searches has error at most $\frac{1}{20n}$.
\end{proof}
  
Theorem~\ref{thm:adj-dir-spanningtree} implies that \textsc{Strong
Connectivity}, too, can be solved using an expected number of queries
in $O(n^{3/2}\sqrt{\log n})$ in the adjacency matrix model.  In fact
the previous algorithm can be used for the matrix model with
$m=n(n-1)$.  We first check that some fixed vertex $v_0$ can reach any
other vertex $u$, producing some spanning tree rooted at $v_0$.  We
then check that all vertices can reach $v_0$ by repeating the previous
step on the transposed adjacency matrix.  The two stages produces a
set at most $2(n-1)$ edges that proves the graph is strongly
connected.

This is not possible in array model, since we store for any vertex $u$
only the neighbors at which edges are pointing, and there is no easy
access to vertices which are connected with a directed edge \emph{to}
$u$.  The following theorem circumvents this obstacle.

\begin{theorem}\label{thm:arr-dir-spanningtree}
  Given a directed graph $G$ in the array model and a vertex $v_0$,
  the algorithm below uses $O(\sqrt{nm \log n})$ queries and outputs
  an edge set $E' \subseteq E$ covering $v_0$.  If $G$ is strongly
  connected, then with probability at least $\frac{1}{4}$ $E'$ is
  strongly connected.
\end{theorem}

\begin{proof}
  In a first stage we use the previous algorithm to construct a
  directed depth first spanning tree $A \subseteq E$ rooted in $v_0$.
  Assume vertexes to be named according to the order they are added to
  $T$.
    
  Then in a second stage we search for every vertex $v_i\in V$, the
  neighbor $v_j$ with smallest index.  The result is a set of backward
  edges $B\subseteq E$.  We claim that the graph $G(V,E)$ is strongly
  connected iff its subgraph $G'(V,A \cup B)$ is strongly connected.
    
    Clearly if $G'$ is strongly connected then so is $G$ since $A\cup
    B\subseteq E$.  Therefore to show the converse assume $G$ strongly
    connected.  For a proof by contradiction let $v_i$ be the vertex
    with smallest index, which is not connected to $v_0$ in $G'$.
    However by assumption there is a path in $G$ {from} $v_i$ to $v_0$.
    Let $(v_l, v_{l'})$ be its first edge with $l\ge i$ and $l'<i$.
    We use the following property of depth first search.
    \begin{quote}
      \begin{lemma}
        \label{lem:prof}
        Let $v_l$ and $v_{l'}$ be two vertexes in the graph $G$ with
        $l<l'$. If there is a path {from} $v_l$ to $v_{l'}$ in $G(V,E)$
        then $v_{l'}$ is in the subtree of $G''(V,A)$ with root $v_l$.
      \end{lemma}
    \end{quote}
    
    Therefore we can replace in the original path the portion {from}
    $v_i$ to $v_l$ by a path only using edges {from} $A$.  Let $v_{l''}$
    be the neighbor of $v_l$ with smallest index.  Clearly $l''\leq l'
    <i$.  By the choice of $v_i$, there exists a path {from} $v_{l''}$
    to $v_0$ in $G'$.  Together this gives a path {from} $v_i$ to $v_0$
    in $G'$ contradicting the assumption and therefore concluding the
    correctness of the algorithm.
    
    Now we analyze the complexity.  During the first stage, set $A$ is
    computed in time $O(\sqrt{n^{3/2}\log n})$.  The second stage can
    be done with $O(\sqrt{nm})$ queries using the minima finding for
    the mapping {from} an edge number in $[1,m]$ to the source-target
    vertex pair.  Both stages can be made succeed with probability at
    least $7/8$.
  \end{proof}

\subsection{The matrix model}

Here all we need is to construct an oriented tree, rooted in some
vertex $v_0$, and it not need to be depth-first.  We want this tree to
cover all vertices reachable by $v_0$.  There is a tricky method for
constructing such a tree with bounded error, without a log-factor in
the running time as the previous algorithm.

The idea is to classify vertices covered by the current tree, into
sets $T_0,\ldots,T_q$ such that the \emph{confidence} that vertices
from $T_i$ have no new neighbors is increasing with $i$.  Whenever a
search of an edge $(u,v)$ with $u\in R$ and $v\not\in
T_0\cup\ldots\cup T_q$ is successful, for some subset $R\subseteq
T_i$, the vertices $R$ and $u$ will be moved into $T_0$, otherwise,
$R$ will be moved into $T_{i+1}$.  We make it formal now.

\begin{enumerate}
\item Let $S$ be the tree consisting of the single vertex $v_0$.  \\
  Let be the partitioning of the vertex set covered by $S$ into
  $T_0=\{v_0\}$ and $T_1=\ldots=T_q=\{\}$, for $q=\lfloor
  \log_2(n)\rfloor+1$.
\item While there is a set $T_i$ with $|T_i|\geq 2^i$ do
  \begin{enumerate}
    \item
      Let $i$ be the smallest index such that $|T_i|\geq 2^i$.
    \item
      If $|T_i|<2^{i+1}$, $R=T_i$ otherwise $R$ is an arbitrary subset of
      $T_i$ with $|R|=2^i$.
    \item
      Remove $R$ from $T_i$.
    \item Search for an edge $(u,v)$ with $u\in R$ and $v\not\in S$ in
      search space of size $O(2^i n)$ with the version of the quantum
      search procedure which uses $O(2^{3i/4}\sqrt{n})$ queries and
      find a solution with probability $1-1/2^{\sqrt{2^{i+2}}}$
      provided such an edge exists.  
    \item
      If the search was successful, add $(u,v)$ to $S$ and $R\cup
      \{v_0\}$ to $T_0$, otherwise add $R$ to $T_{i+1}$.
  \end{enumerate}
  \item
    Output $S$.
\end{enumerate}

Now we show some properties of the algorithm.  For convenience we
define $t_j=|T_0|+|T_1|+\ldots+|T_j|$.
\begin{lemma}
  At the beginning and the end of an iteration we have the following
  invariant.  Let $k$ be the smallest index and $\ell$ be the largest
  index of a non-empty set $T_j$.  Then
\begin{linenomath}\begin{align}
        &|T_{\ell}| \geq 2^{\ell-1}        \label{invar:last}\\
        &\forall k\leq j < \ell : t_j\geq 2^j \label{invar:before}
\end{align}\end{linenomath}
\end{lemma}
\begin{proof}
  by induction on the iterations of the
  algorithm.  Initially, when $T_0$ is the unique non-empty set, the
  claim holds.
  
  Assume the claim holds before an iteration.  
  
  First observe that by the induction assumption~(\ref{invar:before})
  if $k<\ell$ then $|T_k|\geq 2^k$, and so the index chosen by the
  algorithm is always $i=k$.  Now if the search was successful, we
  have already $t_0\geq 2^k$, and therefore $t_j\geq 2^k$ for all
  $0\leq j \leq k$ and for all $j>k$, $t_j$ increased by one,
  preserving condition~(\ref{invar:before}).  If the search was not
  successful, then by the choice of $R$, after the iteration either
  $t_k=0$ or $t_k\geq 2^k$.  For all values $j>k$, $t_j$ is not
  modified, preserving condition~(\ref{invar:before}).

  Condition~(\ref{invar:last}) is preserved because after decreasing
  every set $T_j$ satisfies either $|T_j|=0$ or $|T_j|\geq 2^j$, and
  because whenever a set $T_j$ becomes non-empty, it contains at least
  $2^{j-1}$ elements.
\end{proof}

As a consequence, when the algorithm stops, there is a unique
non-empty set $T_i$ and moreover $2^{i-1}\leq |T_i| < 2^i$.  Also
since at most $n-1$ searches can be successful, the algorithm stops
after $O(n\log n)$ iterations.

\begin{lemma}
  When the algorithm stops, $S$ covers all vertices reachable by $v_0$
  with probability at least $2/3$.
\end{lemma}
\begin{proof}
  Suppose the algorithm failed so there is an edge $(u,v)$ with $u\in
  S$ and $v\not\in S$ which was not found for each call to the search
  procedure with a set $T_j$ containing $u$.  The probability of this
  event is the product of the failure probability over all calls.  It
  is roughly upper bounded by the failure probability of the last
  call, which is at most $p_i:=2^i/2^{\sqrt{2^{i+2}}}$, where $i=|S|$.
  Let $q_i$ be the probability that the algorithm outputs a vertex set
  $S$ with $|S|=i$.  Then the failure probability is upper bounded by
\begin{linenomath}\[
        \sum_{i=0}^n q_ip_i \leq \max_{i=0}^n p_i = p_1 \leq 1/3.
\]\end{linenomath}
\end{proof}

Now we analyze the complexity of the algorithm.
\begin{lemma}
  The expected number of queries done by the algorithm is $O(|S|\sqrt
  n)$.
\end{lemma}
\begin{proof}
  To analyze the total number of queries done by the search
  procedures, we group the calls of the search procedures into
  sequences of unsuccessful searches ending with a success, plus the
  last sequence of unsuccessful searches.
  
  For the first case, let $(u,v)$ be an arbitrary edge found by the
  algorithm.  Then the probability that it was found when $u\in T_i$
  is upper bounded by the probability that it was not found when $u\in
  T_{i-1}$, which is $1/2^{\sqrt{2^i}}$.  The cost of this search and
  the $i-1$ unsuccessful searches over sets $R$ containing $u$ is
  order of
  \begin{linenomath}\[
    \sum_{j=0}^i 2^{3i/4} \sqrt{n} = O(2^{3i/4}\sqrt{n}).  
  \]\end{linenomath}
  Therefore the expected cost of finding $(u,v)$ is at most
  \begin{linenomath}\[
    \sum_{i=0}^q \frac{ 2^{3i/4} }{2^{\sqrt{2^i}}}\sqrt{n} = O(\sqrt{n}).
  \]\end{linenomath}
  
To complete the analysis we upper bound the total work of all the
$O(\log n)$ unsuccessful searches which were made after the last
successful search.  Let $i$ be such that $2^{i-1}\leq |S| < 2^i$.
There where at most $2^i/2^j$ searches for sets $R$ with $|R|=2^j$.
Therefore the total work is order of
  \begin{linenomath}\[
    \sum_{i=0}^i 2^{i-j} 2^{3j/4}\sqrt{n} = O(|S|\sqrt{n}).
  \]\end{linenomath}
This concludes the proof.
\end{proof}

\section{Single source shortest paths}
\label{sec:sssp}

Let $G$ be a directed graph with non-negative edge weights and a fixed
vertex $v_0$.  We want to compute for every vertex $v$ a shortest path
{from} $v_0$ to $v$.  It may happen that it is not unique.  Using for
example the lexicographical ordering on vertex sequences, we choose to
compute a single canonical shortest path.  {From} now on assume that
different paths have different lengths.  As a result, the union over
all vertices $v$ of the shortest paths {from} $v_0$ to $v$ is a
\emph{shortest path tree}.  Let $\nu(u,v)$ be the weight of edge
$(u,v)$ and $\opti{v}$ the shortest path length {from} $v_0$ to $v$.

Classically \textsc{Single Source Shortest Path} may be solved by
Dijkstra's algorithm.  It maintains a subtree $\tree$ with the
``shortest path subtree'' invariant: for any vertex $v\in\tree$, the
shortest path {from} $v_0$ to $v$ uses only vertices {from} $\tree$.
An edge $(u,v)$ is called a \emph{border edge} (of $\tree$) if
$u\in\tree$ and $v\not\in\tree$, and $u$ is called the source vertex,
$v$ the target vertex.  The \emph{cost} of $(u,v)$ is
$\opti{u}+\nu(u,v)$.  Dijkstra's algorithm starts with $\tree=\{v_0\}$
and iteratively adds the cheapest border edge to it.

Our improvement lays in the selection of the cheapest border edge.  We
give the algorithm for the array model.  Setting $m=n^2$ implies the
required bound for the matrix model.

\begin{theorem}
The bounded error query complexity of single source shortest path in
the array model is $O(\sqrt{nm}\log^{3/2} n)$.
\end{theorem}
\begin{proof}
As in Dijkstra's algorithm we construct iteratively a tree $T$, such that
for every vertex $v\in T$, the shortest path {from} $v_0$ to $v$ is in
$T$. We also maintain a partition of the vertices covered by $T$, into a
set sequence. Its length is denoted by $l$.
\begin{enumerate}
\item $T=\{v_0\}$, $l=1$, $P_1=\{v_0\}$
\item Repeat until $T$ covers the graph
\begin{enumerate}
\item For $P_l$ compute up to $|P_l|$ cheapest border edges with
  disjoint target vertices. For this purpose set $N=\sum_{v\in P_l}
  d(v)$, and number all edges with source in $P_l$ {from} $1$ to $N$.
  Define the functions $f:[N]\rightarrow \mathbb N^*$ and
  $g:[N]\rightarrow V$, where $g(i)$ is target vertex of the \nth{i}
  edge and $f(i)$ is its weight if $g(i)\not\in T$ and $\infty$
  otherwise.  Apply the algorithm of section~\ref{sec:minima} on $f$ and
  $g$ with $d=|P_l|$ to find the $d$ lowest cost edges with distinct
  target vertices.  Let $A_l$ be the resulting edge set.
\item Let $(u,v)$ be the minimal weighted edge of $A_1 \cup\ldots\cup
A_l$ with 
$v\not\in P_1\cup\ldots P_l$.
Set  $T=T\cup \{(u,v)\}$, $P_{l+1}=\{v\}$ and $l=l+1$.
\item
As long as $l\geq 2$ and $|P_{l-1}|=|P_l|$, merge $P_l$ into $P_{l-1}$,
and set $l=l-1$.
\end{enumerate}
\end{enumerate}

All steps but 2(b) constructed a vertex set sequence $P_1,\ldots,P_l$,
the cardinality of each being a power of 2, and of strictly decreasing
sizes.  Figure~\ref{fig:decomposition} shows an example of this
partitioning of the vertices in $T$.

\begin{figure}[htb]
\centerline{\input{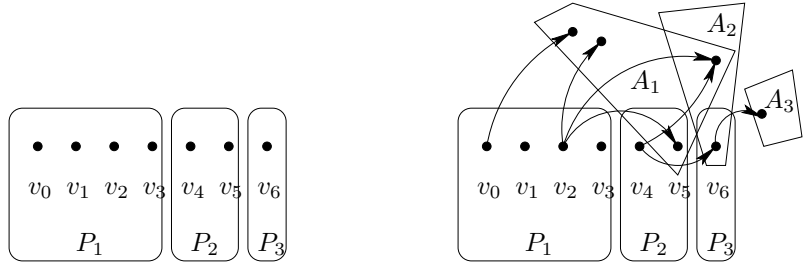}}
\caption{Left: example of the set decomposition for $|T|=7$ into powers of 2. 
  Right: example of corresponding edge sets. The closest border
  edge of $T$ belongs to one of $A_1,A_2,A_3$.}
\label{fig:decomposition}
\end{figure}

Therefore each set $P_i$ is strictly larger than the union of all
the following sets, since $\sum_{i=0}^{k-1}2^i=2^k-1$. 
If $A_i$ contained $|P_i|$ edges, than at least one
of them has its target vertex outside of $P_1\ldots,P_l$. Let $(u,v)$ be
the cheapest border edge of $T$. Let $P_i$ be the vertex set containing
$u$. Then $A_i$ must contain this edge, and step 2(b) selects it.

Only step 2(a) generates queries to the graph.  What is the total
number of queries related to sets $P_i$ of some size $s$?  There are
at most $n/s$ sets of this size $s$.  Therefore total work is order of
$\sum_{j=1}^{n/s} \sqrt{s m_j}$, where $m_j$ is the number of edges
with source in the \nth{j} vertex set.  We have $\sum_{j=1}^{n/s} m_j
= m$.  This worst case is when $m_i=sm/n$. In that case the total work
is $O(\sqrt{nm})$ for the fixed size $s$.  There are $\log n$
different set sizes in the algorithm.  Each of the $O(n\log n)$
queries to the minimum finding procedures should succeed with
probability $1-1/2n\log n$ at least This introduces a $O(\log n)$
factor, see Remark~\ref{remark:min}, and we obtain the claimed
complexity.
\end{proof}

\section{The lower bounds}
\label{sec:lower}

\begin{theorem}
  The problems \textsc{Find $d$ smallest values of a function},
  \textsc{Find $d$ elements of different type} and \textsc{Find $d$
  smallest values of different type} require $\Omega(\sqrt{dN})$
  queries.
\label{lem:allmin}
\end{theorem}
  \begin{proof} For even $k$ and odd $d$ we consider $d\times k$
    boolean matrices with a single $0$ in every row.  It is encoded by
    a function $f:[N]\rightarrow \{0,1\}$ with $N=kd$, such that for
    every $i\in[d],j\in[k]$, $f(id+j)$ is the entry in row $i$ and
    column $j$.  Let function $g:[N]\rightarrow [d+1]$ be such that
    $g(id+j)$ maps to $i$ if $f(id+j)=0$ and to $d$ otherwise.

    So the problems of \textsc{Finding $d+1$ smallest values of a
    function} or \textsc{$d+1$ elements of different type} or
    \textsc{$d+1$ smallest values of different type} are all
    equivalent to finding the positions of the $d$ zeroes in the
    matrix.
        
    Let $X$ be the set of matrices such that exactly $\lfloor
    d/2\rfloor$ rows have their $0$ in the first $k/2$ columns.  And
    let $Y$ be the set of matrices such that this number is exactly
    $\lceil d/2 \rceil$.  We show a lower bound for
    distinguishing $X$ and $Y$.  We say that matrix $A\in X$ is in
    relation with $B\in Y$ iff both matrices differ at exactly two
    entries.  It follows that there are indices $i\in[d], 0\leq
    j<k/2\leq j'<k$ with $A_{ij}=B_{ij'}=1$ and $A_{ij'}=B_{ij}=0$.
    The following example illustrates this definition.

\begin{linenomath}\[
      A=
      \left(\begin{array}{c@{}} \overbrace{011}^{k/2} \\{111}\\ {111} \end{array}
            \begin{array}{@{}c} \overbrace{111}^{k/2} \\{011}\\ {101} \end{array}\right)
      \:\:
      B=
      \left(\begin{array}{c@{}} \overbrace{011}^{k/2} \\{110}\\ {111} \end{array}
            \begin{array}{@{}c} \overbrace{111}^{k/2} \\{111}\\ {101} \end{array}\right)
\]\end{linenomath}
  
    Then the number of matrices which are in relation with a fixed matrix
    is at least $m=m'=\lceil d/2\rceil k/2$.  For fixed $i,j$
    the number of matrices in relation with $M$ and differing at
    $i,j$ is $k/2$ if $M_{ij}=0$ and $1$ if $M_{ij}=1$.  So
    $l_{\max}=k/2$ and $\Omega(\sqrt{mm'/l_{\max}})$ gives the required
    lower bound.
  \end{proof}

Now, we give a simple lower bound for \textsc{Connectivity} (and
\textsc{Strong Connectivity}) in the list model, by a reduction {from}
\textsc{Parity}.  As we recently found out, this reduction has first
been used by Henzinger and Fredman for the on-line connectivity
problem~\cite{HF98}.  We show later how to improve this construction.
  \begin{lemma}
    \textsc{Strong Connectivity} requires $\Omega(n)$ queries in the
    array model.
    \label{lem:parity}
  \end{lemma}
  \begin{proof}
    We use a straightforward reduction {from} \textsc{Parity}.

    \begin{figure}[ht]
      \centerline{\input{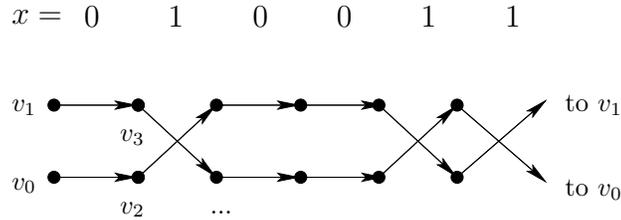}}
      \caption{A standard reduction {from} parity}
      \label{fig:parity}
    \end{figure}
    
    Let $ x\in \{0,1\}^{p} $ be an instance of the parity problem. We
    construct a permutation $ f $ on $ V=\{v_0,\ldots,v_{2p-1}\} $
    which has exactly 1 or 2 cycles depending the parity of $ x $. For
    any $ i\in [p] $ and the bit $b=x_i$, we define $
    f(v_{2i})=v_{2i+2+b}$ and $ f(v_{2i+1})=v_{2i+3-b}$, where addition
    is modulo $2p$.  See figure~\ref{fig:parity}.  The graph defined
    by $ f $ has 2 levels and $ p $ columns, each corresponding to a
    bit of $ x $. A directed walk starting at vertex $ v_0 $ will go
    {from} left to right, changing level whenever the corresponding bit
    in $ x $ is 1. So when $ x $ is even the walk returns to $ v_0 $
    while having explored only half of the graph, otherwise it returns
    to $ v_1 $ connecting {from} there again to $ v_0 $ by $ p $ more
    steps.  Since the query complexity of \textsc{Parity} is
    $\Omega(n)$ --- see for example \cite{FGGS} --- this concludes the
    proof.
  \end{proof}
  
  The same technique can be extended to the undirected case.

\begin{corollary}
  \textsc{Connectivity} requires $\Omega(n)$ queries in the array model.
\end{corollary}

We improve the lower bound in Lemma~\ref{lem:parity} by changing the
construction slightly.

  \begin{theorem}
    \textsc{Strong Connectivity} requires $ \Omega (\sqrt{nm}) $ queries
    in the array model.\label{thm:hyper}
  \end{theorem}
  \begin{proof}
    Let $m$ be such that $m=kn$ for some integer $k$.  We construct
    the lower bound for graph with regular out-degree $k$.
    
    We use a similar construction as for Lemma~\ref{lem:parity}, but
    now for every vertex the $k-1$ additional edges are redirected
    back to an origin.  We would like to connect them back to a fixed
    vertex $u_0$, but this would generate multiple edges and we want
    the proof work for simple graphs.  Therefore we connect them back
    to a $k$-clique which then is connected to $u_0$.  See
    figure~\ref{fig:origin}.  Let vertex set $V=\{v_{0},\ldots
    ,v_{2p-1},u_{0},\ldots ,u_{k-1}\} $ for some integer $p$.  In the
    list model, the edges are defined by a function $f:V\times
    [k]\rightarrow V $.  We consider only functions with the
    following restrictions:
    
    For every $i\in [k]$ we have $f(u_i,0)=v_0$ and for
    $j\in\{1,\ldots,k-1\}$ $f(u_i,j)=u_{i+j}$, where addition is
    modulo $k$.
    
    For every $i\in [p]$ there exist $j_0, j_1\in [k]$ and a bit $b$
    such that $f(v_{2i},j_0)=v_{2i+2+b}$ and
    $f(v_{2i+1},j_1)=v_{2i+3-b}$, where addition is modulo $2p$ this
    time.  We call these edges the \emph{forward edges}.  The
    \emph{backward edges} are for all $j\in[k]$ $f(v_{2i},j)=u_j$
    whenever $j\neq j_0$ and $f(v_{2i+1},j)=u_j$ whenever $j\neq j_1$.

    \begin{figure}[ht]
      \centerline{\input{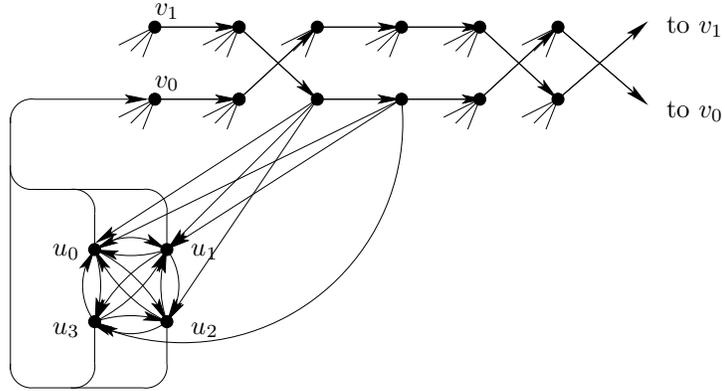}}
      \caption{A strongly connected graph}
      \label{fig:origin}
    \end{figure}
    
    Now all the vertices are connected to the $k$-clique, the clique
    is connected to $v_{0}$, and the graph is strongly connected if
    and only if there is a path {from} $v_{0}$ to $v_{1}$.

    Let $X$ be the set of functions which define a strongly connected
    graph, and $Y$ the set of functions which do not.  Function $ f\in
    X $ is in relation with $g\in Y$ if there are numbers $
    i\in[p],j_{0},j_{1},h_{0},h_1\in[k]$ with $j_0\neq h_0$, $j_1\neq
    h_1$ such that the only places where $ f $ and $ g $ differ are
    \begin{linenomath}\begin{eqnarray*}
      g(v_{2i},h_{0})=f(v_{2i+1},j_{1}) &  
                    & g(v_{2i+1},h_{1})=f(v_{2i},j_{0})         \\
      g(v_{2i},j_{0})=u_{j_0}&
                    & g(v_{2i+1},j_{1})=u_{j_1}                 \\
      f(v_{2i},h_{0})=u_{h_0} &  
                    & f(v_{2i+1},h_{1})=u_{h_1}
    \end{eqnarray*}\end{linenomath}
    Informally $f$ and $g$ are in relation if there is a level, where
    the forward edges are exchanged between a \emph{parallel} and
    \emph{crossing} configuration and in addition the edge labels are
    changed.

    Then $m=m'=O(nk^{2})$, $p\in O(n)$ for the number of levels and
    $(k-1)^2$ for the number of possible forward edge labels.  We also
    have $l_{f,(v,j)}=k-1$ if $f(v,j)\in \{u_{0},\ldots ,u_{k-1}\} $ and
    $l_{f,(v,j)}=(k-1)^2$ otherwise.  The value $l'_{g,(v,j)}$ is the
    same.  Since only one of $f(v,j)$, $g(v,j)$ can be in
    $\{u_{0},\ldots ,u_{k-1}\} $ we have $ l_{\max }=O(k^3)$ and the
    lower bound follows.
  \end{proof}
  
  For the matrix model, there is a much simpler lower bound which
  works even for undirected graphs.

  \begin{theorem}                               \label{thm:2cycles}
    \textsc{Connectivity} requires $ \Omega (n^{3/2}) $ queries in the
    matrix model.
  \end{theorem}
  \begin{proof}
    We use Ambainis' method for the following special problem. You are
    given a symmetric matrix $ M\in \{0,1\}^{n\times n}$ with the
    promise that it is the adjacency matrix of a graph with exactly
    one or two cycles, and have to find out which is the case.
    
    Let $X$ be the set of all adjacency matrices of a unique cycle,
    and $Y$ the set of all adjacency matrices with exactly two cycles
    each of length between $n/3$ and $2n/3$. We define the relation
    $R\subseteq X\times Y$ as $M\;R\;M'$ if there exist $ a,b,c,d\in
    [n] $ such that the only difference between $M$ and $M'$ is that
    $(a,b), (c,d)$ are edges in $M$ but not in $M'$ and $(a,c), (b,d)$
    are edges on $M'$ but not in $M$.  See figure~\ref{fig:cycles}.
    The definition of $Y$ implies that in $M$ the distance {from} $a$ to
    $c$ is between $ n/3 $ and $ 2n/3 $.

    \begin{figure}[ht]
      \centerline{\epsfig{file=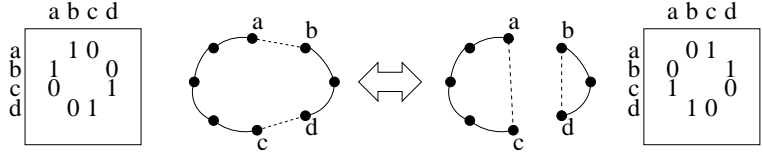,width=10cm}}
      \caption{Illustration of the relation}
      \label{fig:cycles}
    \end{figure}

    Then $m=O(n^{2})$ since there are $n-1$ choices for the first edge
    and $n/3$ choices for the second edge.  Also $m'=O(n^2)$ since
    {from} each cycle one edge must be picked, and cycle length is at
    least $n/3$.
    
    We have $l_{M,(i,j)}=4$ if $M_{i,j}=0$ since in $M'$ we have the
    additional edge $(i,j)$ and the endpoints of the second edge must
    be neighbors of $i$ and $j$ respectively.  Moreover
    $l_{M,(i,j)}=O(n)$ if $M_{i,j}=1$ since then $(i,j)$ is one of the
    edges to be removed and there remains $n/3$ choices for the second
    edge.
    
    The values $l'_{M',(i,j)}$ are similar, so in the product one
    factor will always be constant while the other is linear giving
    $l_{M,(i,j)}l'_{M',(i,j)}=O(n)$ and the theorem follows.
  \end{proof}
  
  We give a lower bound for both minimum spanning tree and
  single-source shortest paths.
\begin{theorem}
  Finding minimum spanning tree and single source shortest paths require
  $\Omega(\sqrt{nm})$ queries.
\end{theorem}
\begin{proof}
  The proof is a reduction {from} minima finding.  Let $m=k(n+1)$ for
  some integer $k$.  Let $M$ be a matrix with $n$ rows and $k$ columns
  and positive entries.  The lower bound on minima finding, with
  $d=n,N=kn$, shows that $\Omega(\sqrt{kn^2})$ queries are required to
  find the minimum value in every row.

  \begin{figure}[htbp]
    \centering{\input{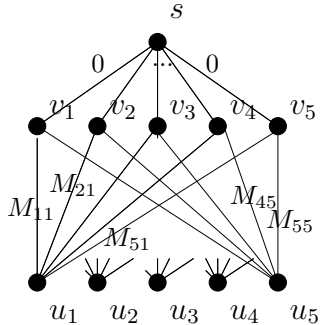}}
    \caption{Reduction {from} finding minima to minimal spanning tree}
    \label{fig:red_mst}
  \end{figure}
  
  We construct a weighted graph $G$ {from} $M$, like this: The
  vertices are $V(G)=\{s,v_1,\ldots,v_k,u_1,\ldots,u_{n}\}$.  The
  edges are all $(s,v_i)$ with weight $0$ and all $(v_i,u_j)$ with
  weight $M_{ji}$.  (See figure~\ref{fig:red_mst}.) Then clearly a
  minimum spanning tree contains the $0$-weight edges connecting $s$
  to all vertices $v_i$.  And every vertex $u_j$ will be connected to
  the rest of the graph only with the minimal weighted edge.
\end{proof}

\section*{Acknowledgments}
  
For helpful discussions or comments we are grateful to Miklos Santha,
Katalin Friedl, Oded Regev, Ronald de Wolf and Andris Ambainis.

\end{document}